\documentclass[prb,twocolumn,aps,showpacs,showkeys]{revtex4}
\usepackage{graphicx}

\begin{document}
\title{
Driven linear modes: Analytical solutions for finite discrete systems
}
\author{N. Lazarides$^{1,2}$ and G. P. Tsironis$^{1}$
}
\affiliation{
Department of Physics, University of Crete, and
Institute of Electronic Structure and Laser, Foundation for Research
and Technology-Hellas,
P. O. Box 2208, 71003 Heraklion, Greece \\
$^{2}$Department of Electrical Engineering, Technological
Educational Institute of Crete, P. O. Box 140, Stavromenos, 71500,
Heraklion, Crete,
Greece
}

\date{\today}
\begin{abstract}
  We have obtained exact analytical expressions in closed form,
  for the linear modes excited in finite and discrete systems 
  that are driven by a spatially homogeneous alternating field.
  Those modes are extended for frequencies within the linear frequency 
  band while they are either end-localized or end-avoided for frequencies
  outside the linear frequency band.
  The analytical solutions are resonant at particular frequencies,
  which compose the frequency dispersion relation of the finite system.

\end{abstract}


\maketitle

{\em Introduction.-} 
The calculation of the linear modes which can be excited 
in finite and discrete systems 
that are driven by a spatially homogeneous alternating (AC) field
is of great interest. 
That problem arises in the low-amplitude limit of several model equations 
of physical systems,
which are comprised of elements that are either directly or indirectly 
coupled (i.e., the so called magnetoinductive systems).
In that limit, it is a common practice to omit the nonlinear terms
ending up with a linear non-autonomous system of equations that 
accept linear wave solutions called in the following driven linear 
modes (DLM).
Such a linearized problem may arise, for example, from the AC driven 
Frenkel-Kontorova
model \cite{Bonilla,Floria,Braun}, which has been widely used to model 
rf-driven  parallel arrays of Josephson junctions (JJs) \cite{Mazo,Zueco,Falo},
or from more general models of AC driven anharmonic lattices 
with realistic potentials \cite{Rossler1995,Rossler1996}.
It may also arise from models of inductively coupled AC driven
rf-SQUID arrays \cite{Lazarides2007,Lazarides2008,Tsironis},
from models of inductively coupled intrinsic Josephson junctions
supplied by AC current
\cite{Sakai1993,Pedersen1998,Pedersen2005,Madsen},
and also from nonlinear magnetic metamaterial models that are comprised
of split-ring resonators placed in an AC magnetic field  
\cite{Lazarides2006,Kourakis,Eleftheriou}.
In the latter case, those linear waves are known as (linear) magnetoinductive 
waves \cite{Shamonina2002,Shamonina2008}, and they have phonon-like 
dispersion curves \cite{Sydoruk2005} and many prospects for device applications
\cite{Syms2006a,Shamonina2004,Syms2006b}.
In the present work we present analytical solutions that we have obtained
for DLMs in finite and discrete systems, with frequencies either in the 
linear wave band (LWB) obtained in the usuall way, or outside that band.  
We find that the former ones are extended modes as it is expected, while the
latter ones are either end-localized or end-avoided modes.
The results are illustrated using as a paradigmatic example the driven
FK chain \cite{Floria}, which in a standard normalization form reads
\begin{equation}
  \label{99}
  \ddot{q}_i +\alpha \dot q_i + \frac{1}{2\pi} \sin(2\pi q_i) =
  C ( q_{i-1} -2 q_i + q_{i+1} ) +f_{ac} \sin(\omega t) ,   
\end{equation}
where $q_i$ is the $i$th 'coordinate' whose interpretations depends on the 
particular system to which Eq. (\ref{99}) is related,
$\alpha$ is the loss coefficient, $C$ is the coupling constant,
$f_{ac}$ is the amplitude of the sinusoidal driver with frequency $\omega$,
and  $i=1,...,N$, with $N$ being the total number of elements in the chain.

{\em The linearized problem.-}
Linearization of  Eq. (\ref{99}) gives 
\begin{equation}
 \label{98}
  \ddot{q}_i +q_i = 
    C ( q_{i-1} -2 q_i + q_{i+1} ) +f_{ac} \sin(\omega t) ,   
\end{equation}
where the losses have been omitted for simplicity. The earlier equation 
accepts solutions of the form $q_i (t) = Q_i\, \sin(\omega t)$, 
where $Q_i$ is the time-independent mode amplitude at site $i$.
By substitution of $q_i (t)$ into Eq. (\ref{98}) we obtain a system of 
stationary equations 
\begin{equation}
  \label{1}
      s\, Q_{i-1} +Q_i + s\, Q_{i+1} = \kappa
\end{equation}
where $s$ and $\kappa$ are real parameters given by
\begin{equation}
 \label{1a}
   s=\frac{C}{\omega^2 -(1+2C)}, \qquad \kappa =\frac{f_{ac}}{(1+2C)-\omega^2} .
\end{equation}
In general, $s$ and $\kappa$ depend on the parameters of the particular model.
Since we consider finite systems, Eq. (\ref{1}) should be implemented with 
open-ended boundary conditions, i.e., $ Q_0 = Q_{N+1} = 0$,
to account for the termination of the structure at both ends.
The formal solution of Eq. (\ref{1}) can be written as 
\begin{equation}
  \label{3}
    {\bf Q} = \kappa\, \hat{\bf S}^{-1} {\bf 1} ,
\end{equation}
where  ${\bf Q}$ and ${\bf 1}$ are $N-$dimensional vectors  with
componets $Q_i$ and $1$, respectively,
and $\hat{\bf S}^{-1}$ is the inverse of the $N\times N$  coupling matrix
$\hat{\bf S}$. The latter 
is a real, symmetric tridiagonal matrix that has diagonal elements 
equal to unity, while all the other non-zero elements are equal to $s$.
The need to find the inverse of tridiagonal matrices like $\hat{\bf S}$
arises in many scientific and engineering applications.
Recently,  Huang and McColl \cite{Huang}
related the inversion of a general trigiagonal matrix to second order 
linear recurrences, and they provided a set of very simple analytical formulae
for the elements of the inverse matrix.
Those formulae lead immediately to closed forms for certain, relatively
simple trigiagonal matrices,
like $\hat{\bf S}$.

{\em Analytical solutions.-}
Following Huang and McColl \cite{Huang} we can show that the elements of 
$\hat{\bf S}^{-1}$ are given by ($i \geq j$)
\begin{equation}
  \label{4}
    \left(\hat{\bf S}^{-1} \right)_{ij}=
     \mu \frac{\sinh(j \theta) \, \sinh[(N-i+1)\theta]}
     {\sinh \theta \, \sinh[(N+1)\theta]} , 
\end{equation}
for $-1/2 < s < 1/2$, where 
\begin{equation}
  \label{5}
    \mu = \frac{1}{|s|} \left( -\frac{|s|}{s} \right)^{(i-j)}, 
   \qquad 
   \theta =\ln \frac{1 +\sqrt{1 -(2 s)^2}}{2|s|} ,
\end{equation}    
and 
\begin{equation}
  \label{6}
    \left(\hat{\bf S}^{-1} \right)_{ij}=
     \mu \frac{\sin(j \theta') \, \sin[(N-i+1)\theta']}
     {\sin \theta' \, \sin[(N+1)\theta']} , 
\end{equation}
for $s > 1/2$ or $s < -1/2$, where 
\begin{equation}
  \label{5.1}
    \mu = \frac{1}{|s|} \left( -\frac{|s|}{s} \right)^{(i-j)}, 
   \qquad 
   \theta' =\cos^{-1} \left( \frac{1}{2|s|} \right) .
\end{equation}    
The elements of $\hat{\bf S}^{-1}$ for $i<j$ are given by the same 
expressions as in Eqs. (\ref{4}) and (\ref{7}) by simply
interchanging the indices $i$ and $j$ (due to the symmetry).
Although the above expressions are valid for any $N$, in the following 
we assume that $N$ is even.

For obtaining the solution $\bf Q$,
we first write Eq. (\ref{1}) as 
\begin{equation}
  \label{7}
    Q_i = \kappa \sum_{j=1}^N \left( \hat{\bf S}^{-1} \right)_{ij} .
\end{equation}    
It turns out that the sum in the right-hand-side of the earlier 
equation can be be calculated in closed form, using in an appropriate
way the summation formulae \cite{Gradshteyn} 
\begin{eqnarray}
\label{8}
  \sum_{j=1}^{N-1} \sinh(j y) =\sinh\left(\frac{N-1}{2}y\right)
    \frac{\sinh\left(\frac{N}{2}y\right)}
       {\sinh\frac{y}{2}} \\
 \label{9}
  \sum_{j=1}^{N-1} \cosh(j y) =\cosh\left(\frac{N-1}{2}y\right)
    \frac{\sinh\left(\frac{N}{2}y\right)}
       {\sinh\frac{y}{2}} -1, 
\end{eqnarray}
for $-1/2 < s < +1/2$, and the formulae \cite{Gradshteyn}
\begin{eqnarray}
\label{10}
  \sum_{j=1}^{N} \sin(j y) =\sin\left(\frac{N+1}{2}y\right)
    \frac{\sinh\left(\frac{N}{2}y\right)}
       {\sinh\frac{y}{2}} \\
 \label{11}
  \sum_{j=1}^{N} \cosh(j y) =\cos\left(\frac{N+1}{2}y\right)
    \frac{\sinh\left(\frac{N}{2}y\right)}
       {\sinh\frac{y}{2}} , 
\end{eqnarray}
for $s > 1/2$ or $s < -1/2$.
Then, using several identities for hyperbolic and trigonometric functions,
we arive, after long and tedious calculations to the solution
\begin{eqnarray}
\label{12}
  Q_{i=odd} =\frac{\kappa}{2 |s| D_H^+} \cosh\left(\frac{i\theta}{2}\right)
                    \sinh\left(\frac{N-i+1}{2}\theta\right)  \\
\label{13}
  Q_{i=even} =\frac{\kappa}{2 |s| D_H^+} \sinh\left(\frac{i\theta}{2}\right)
                    \cosh\left(\frac{N-i+1}{2}\theta\right) ,
\end{eqnarray}
where
\begin{equation}
\label{14}
  D_H^+ = \cosh^2 \left(\frac{\theta}{2}\right) 
       \sinh\left(\frac{N+1}{2}\theta\right) ,
\end{equation}
for $0 < s <+1/2$,
\begin{eqnarray}
\label{15}
  Q_i =\frac{\kappa}{2 |s| D_H^-} \sinh\left(\frac{i\theta}{2}\right)
                    \sinh\left(\frac{N-i+1}{2}\theta\right) , 
\end{eqnarray}
where
\begin{equation}
\label{16}
  D_H^- = \sinh^2 \left(\frac{\theta}{2}\right) 
       \cosh\left(\frac{N+1}{2}\theta\right) ,
\end{equation}
for $-1/2 < s < 0$,
\begin{eqnarray}
\label{17}
  Q_{i=odd} =\frac{\kappa}{2 |s| D_T^+} 
       \cos\left(\frac{i\theta'}{2}\right)
       \sin\left(\frac{N-i+1}{2}\theta' \right)  \\
\label{18}
  Q_{i=even}=\frac{\kappa}{2 |s| D_T^+} 
       \sin\left(\frac{i\theta'}{2}\right)
       \cos\left(\frac{N-i+1}{2}\theta' \right) ,
\end{eqnarray}
where
\begin{equation}
\label{19}
  D_T^+ = \cos^2 \left(\frac{\theta'}{2}\right) 
          \sin\left(\frac{N+1}{2}\theta'\right) ,
\end{equation}
for $s > +1/2$, and
\begin{eqnarray}
\label{20}
  Q_i =\frac{\kappa}{2 |s| D_T^-} 
     \sin\left(\frac{i\theta'}{2}\right)
     \sin\left(\frac{N-i+1}{2}\theta'\right) , 
\end{eqnarray}
where
\begin{equation}
\label{21}
  D_T^- = \sin^2 \left(\frac{\theta'}{2}\right) 
       \cos\left(\frac{N+1}{2}\theta' \right) ,
\end{equation}
for $s < -1/2$.

The case where $s=\pm 1/2$ should be treated separately.
However, the solution of Eqs. (\ref{1}) for that specific value of the 
coupling parameter can be obtained by calculating the limit $s\rightarrow 1/2$
of the expressions Eqs. (\ref{12})-(\ref{13}) and Eqs. (\ref{17})-(\ref{18})
(for $s=+1/2$). We find that the limits of those expressions are the the same, 
so that $Q_i (s=1/2)$ is given by
\begin{eqnarray}
\label{22}
  Q_{i=odd} = \kappa \frac{N-i+1}{N+1} , \\
  Q_{i=even} = \kappa \frac{i}{N+1} .
\end{eqnarray}  
In the same way, by calculating the limit $s\rightarrow -1/2$ of 
the expressions Eq. (\ref{15}) and Eq. (\ref{20}) (for $s=-1/2$)
gives for the $Q_i (s=-1/2)$ the expression
\begin{equation}
\label{23}
  Q_i = \kappa \, i \, (N-i+1) .
\end{equation}
At this point we have given the solutions of Eqs. (\ref{1})
for all real values of the coupling parameter $s$.

{\em Illustrative examples and discussion.-}
For illustration we show in Figs. 1 and 2 (for $s>0$ and $s<0$, respectively)
plots of $Q_i$ as a 
function of $i$ for an FK chain with $N=50$, $C=0.5$, $f_{ac} =0.02$,
and several frequencies.
We should note that those results where checked numerically.
The linear dispersion relation (LDR) for Eq. (\ref{98}) is \cite{Floria}
\begin{equation}
\label{97}
  \omega = \sqrt{ 1 +4C\sin^2\left({k}/{2}\right) },
\end{equation}
where $k$ is the normalized wavenumber. The extends from $\omega_{min}=1$
to $\omega_{max}=\sqrt{1+4C}$, having gaps below and above $\omega_{min}$
and $\omega_{max}$, repsectively. 
It is easy to see that the frequencies within the LWB correspond to
either $s<-1/2$ or $s>+1/2$, for $\omega_{min} < \omega < \omega_s$ and 
$\omega_s < \omega < \omega_{max}$, respectively, 
where $\omega_s =\sqrt{1 +2 C}$.
Several DLMs are shown in Figs. 1a, 1b, and 1c, corresponding to 
$0 <s <+1/2$, $s=+1/2$ and $s > +1/2$, respectively.
We observe that the modes in Fig. 1a, with frequencies above the LWB, 
exhibit weak localization at the end-points of the chain.
The mode shown in Fig. 1b correspond to frequency $\omega=\omega_{max}$,
exactly on the upper bound of the LWB, while those shown in Fig. 1c
are extended stationary solutions, with frequencies in the LWB.
\begin{figure}[!t]
\includegraphics[angle=0, width=0.85\linewidth]{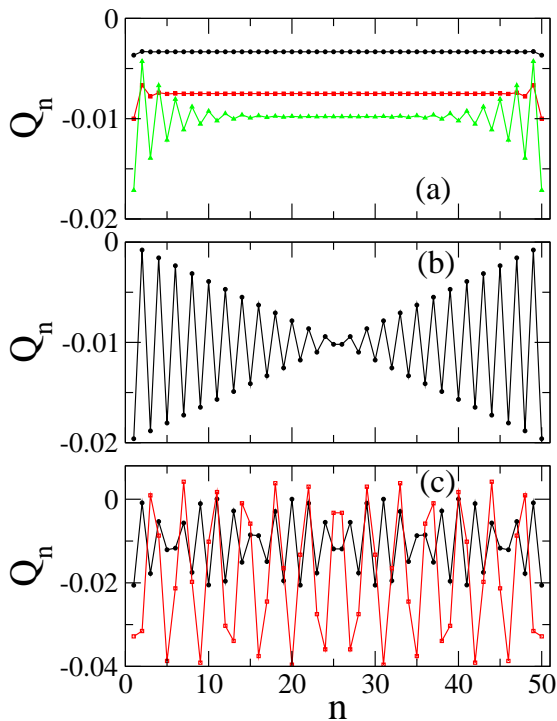}
\caption{(color online)
Driven linear modes for the finite Frenkel-Kontorova chain for 
$C=0.5$, $f_{ac} =0.02$, $N=50$, and $s>0$.
The continuous lines serve as a guide to the eye.
(a) End-localized modes for 
$\omega =2.646$ ($s=0.1$, $\kappa=-0.004$; black circles),
$\omega =1.915$ ($s=0.3$, $\kappa=-0.012$; red squares),
$\omega =1.744$ ($s=0.48$, $\kappa=-0.0192$; green triangles).
(b) Special case of $\omega=\omega_{max}$ ($s=+0.5$, $\kappa=-0.02$). 
(c) Extended modes for 
$\omega =1.717$ ($s=0.52589$, $\kappa=-0.021$;  black circles),
$\omega =1.459$ ($s=3.8781$, $\kappa=-0.0155$;   red squares).
}
\end{figure}
The DLMs shown in Figs. 2a, 2b, and 2c, correspond to 
$0 >s >-1/2$, $s=-1/2$ and $s < -1/2$, respectively.
The modes shown in Fig. 2a, with frequencies below the LWB, 
exhibit end-avoidance.
The mode shown in Fig. 2b correspond to frequency $\omega=\omega_{mim}$,
exactly on the lower bound of the LWB, while those shown in Fig. 2c
are extended stationary solutions.
From Figs. 1 and 2 we observe that the amplitude of the DLMs with 
frequencies in the LWB increases with decreasing frequency.
\begin{figure}[!t]
\includegraphics[angle=0, width=0.8\linewidth]{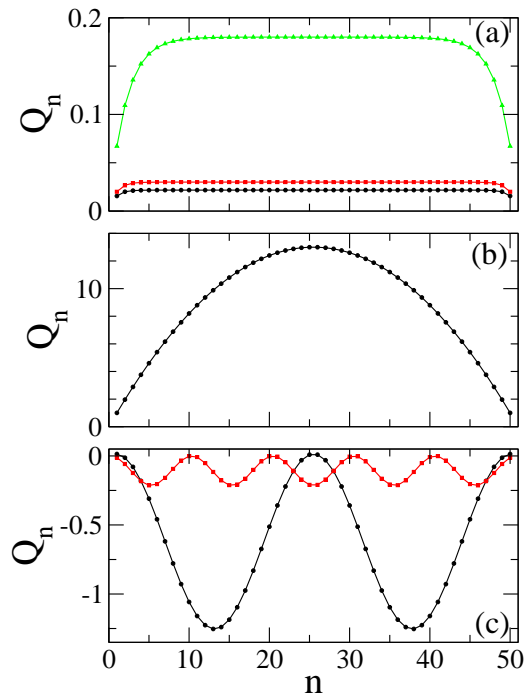}
\caption{(color online)
Driven linear modes for the finite Frenkel-Kontorova chain for 
$C=0.5$, $f_{ac} =0.02$, $N=50$, and $s<0$.
The continuous lines serve as a guide to the eye.
(a) End-avoided modes for 
$\omega =0.2773$ ($s=-0.26$, $\kappa=0.0104$; black circles),
$\omega =0.5774$ ($s=-0.3$, $\kappa=0.012$; red squares),
$\omega =0.9428$ ($s=-0.45$, $\kappa=0.018$; green triangles).
(b) Special case of $\omega=\omega_{min}$ ($s=-0.5$, $\kappa=0.02$). 
(c) Extended modes for 
$\omega =1.016$ ($s=-0.51668$, $\kappa=-0.02067$;  black circles),
$\omega =1.089$ ($s=-0.61493$, $\kappa=-0.02457$;   red squares).
}
\end{figure}

{\em Frequency dispersion and concluding remarks.-}
Note that those amplitudes are determined uniquelly from the parameters
of the system. Even though we have chosen a rather weak driving amplitude
$f_{ac}$, the DLMs close to the lower bound of the LWB attain rather 
large amplitude ($>1$ in some cases in normalized units).
Moreover, from the analytical solutions we infer that the DLMs 
are resonant for particular values of frequencies in the LWB.
The frequencies of those resonances can be obtained by zeroing 
the denominators of the solutions for the DLMs in the Eqs. (\ref{14}),
(\ref{16}), (\ref{19}), and (\ref{21}).
Specifically, by zeroing the denominators $D_T^+$ and $D_T^-$, 
for $s>+1/2$ and $s<-1/2$, respectively, we get for the resonant values
of $s$ the relations
\begin{equation}
  \label{24}
    s_m^R = \left\{2 \cos\left( \frac{2 m \pi}{n+1} \right) \right\}^{-1} ,
\end{equation}
for $s>+1/2$, where the integer $0 < m < (n+1)/4$, while 
\begin{equation}
  \label{25}
    s_m^R = \left\{-2 \cos\left( \frac{(2 m+1) \pi}{n+1} \right) \right\}^{-1} ,
\end{equation}
for $s<-1/2$, where the integer $0 \le m < (n-1)/4$. 
The total number of resonances, which provide the frequency dispersion for
a finite chain, are $n/2$. 
The corresponding resonance frequencies $\omega_m$ are obtained from
the first of Eqs. (\ref{1a}), as
\begin{equation}
  \label{26}
   \omega_m = \sqrt{ \omega_s^2 + ({C}/{s_m^R}) } ,
\end{equation}
with $s_m^R$ given by either by Eq. (\ref{24}) (for $s>+1/2$)
or by Eq. (\ref{25}) (for $s<-1/2$).
Since we ignored the loss term in Eq. (\ref{98}),
the mode amplitudes go to infinity at the resonant frequencies.
We  should note that both in Figs. 1c and 2c, the values of $\omega$ were chosen
in between two neighboring resonances. If losses were taken into account,
the mode amplitudes would still reach considerably high values, 
even for very weak driver. That example questions the validity of any linear
approximation for DLMs at some part of the LWB, which for the particular model
lies between $\omega_s$ and $\omega_{min}$. That situation get worse and worse as
the lower bound is approached, and the large amplitude modes are expected to become
modulationally unstable for moderately high driving amplitudes.

\begin{figure}[!t]
\includegraphics[angle=0, width=0.85\linewidth]{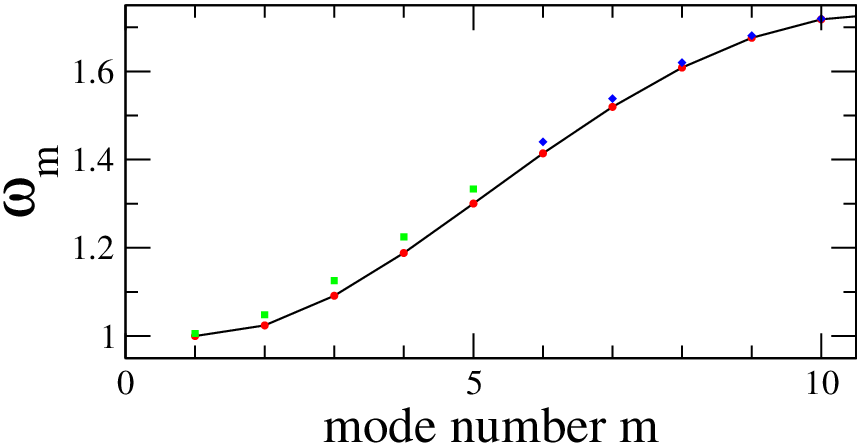}
\caption{(color online)
Frequency dispersion (eigen-frequencies vs. mode number m) for the
finite Frenkel-Kontorova chain with $N=20$, and $C=0.5$. 
The black curve gives the frequency dispersion
for the infinite system while the red circles correspond to the 
eigen-frequencies of the periodic system with $N=20$.
The green squares and the blue rhombii are the eigen-frequencies
of the finite system with open-ended boundary conditions for 
$m$ even and odd, respectively.
}
\end{figure}
We have presented analytical solutions in closed form for the DLMs 
in finite and discrete systems whose elements are coupled with their
nearest neighbors. It should be stressed that these solutions
are generally valid for any such system with the appropriate forms 
$s$ and $\kappa$ on its parameters.
The presented examples, using the driven FK model show that the DLMs
are extended for frequencies in the LWB and end-localized (end-avoided)
for frequencies above (below) the LWB.
Those particular features of DLMs are due to the choise of the 
open-ended boundary conditions, which modifies the local potential that 
an element feels close to the end-points.
The resonance frequencies $\omega_m$, obtained from Eq. (\ref{26})
for all relevant values of $m$,
provide the frequency dispersion for the finite chain,
which differs slightly from that of the periodic system (Fig. 3).
The difference increases with decreasing number of elements.
Preliminary calculations show that the localized states shown here are stable
and, moreover, they may evolve into dissipative surface states 
when the nonlinearity sets in and the losses are included \cite{Lazarides2008b}.

\end{document}